\documentclass[a4paper,english]{lipics-v2019}
\usepackage{nth}
\usepackage{microtype}
\usepackage{breqn}
\usepackage{ifthen}
\usepackage{placeins}
\usepackage{complexity}
\usepackage{nicefrac}
\hideLIPIcs
\nolinenumbers

\newcommand{\old}[1]{{}}

\title{Computing Coordinated Motion Plans for\\ Robot Swarms: The CG:SHOP Challenge 2021}
\titlerunning{Coordinated Motion Planning: CG Challenge 2021}

\author{Sándor P.~Fekete}{Department of Computer Science, TU Braunschweig, Germany}{s.fekete@tu-bs.de}{https://orcid.org/0000-0002-9062-4241}{}
\author{Phillip Keldenich}{Department of Computer Science, TU Braunschweig, Germany}{p.keldenich@tu-bs.de}{https://orcid.org/0000-0002-6677-5090}{}
\author{Dominik Krupke}{Department of Computer Science, TU Braunschweig, Germany}{d.krupke@tu-bs.de}{https://orcid.org/0000-0003-1573-3496}{}
\author{Joseph S.~B.~Mitchell}{Department of Applied Mathematics and Statistics, Stony Brook University, USA}{joseph.mitchell@stonybrook.edu}{https://orcid.org/0000-0002-0152-2279}{}
\authorrunning{S.~P.~Fekete, P.~Keldenich, D.~Krupke, J.~S.~B.~Mitchell}
\Copyright{S.~P.~Fekete, P.~Keldenich, D.~Krupke, J.~S.~B.~Mitchell}



\supplement{}

\funding{Work at TU Braunschweig has been partially supported by the German Research Foundation (DFG), 
project ``Computational Geometry: Solving Hard Optimization Problems'' (CG:SHOP), grant FE407/21-1. J. Mitchell is partially supported by the National Science Foundation (CCF-2007275).} 

\acknowledgements{We thank the members of the CG Challenge Advisory Board for their valuable input: 
William J.~Cook, Andreas Fabri, Michael Kerber, Philipp Kindermann, Kevin Verbeek.}

\EventAcronym{CG:SHOP 2021}
\begin{document}
\maketitle
\begin{abstract}
We give an overview of the 2021 Computational Geometry Challenge,
which targeted the problem of optimally coordinating a set of robots
by computing a family of collision-free trajectories for a set
$S$ of $n$ pixel-shaped objects from a given start configuration
to a desired target configuration.
\end{abstract}

\section{Introduction}

The ``CG:SHOP Challenge'' (Computational Geometry: Solving Hard
Optimization Problems) originated as a workshop at the 2019
Computational Geometry Week (CG Week) in Portland, Oregon in June,
2019.  The goal was to conduct a computational challenge competition
that focused attention on a specific hard geometric optimization
problem, encouraging researchers to devise and implement solution
methods that could be compared scientifically based on how well they
performed on a database of carefully selected and varied instances.
While much of computational
geometry research is theoretical, often seeking provable approximation
algorithms for \NP-hard optimization problems,
the goal of the CG Challenge was to set the metric of success based on
computational results on a specific set of benchmark geometric
instances. The 2019 CG Challenge focused on the problem of computing
minimum-area polygons whose vertices were a given set of points in the
plane.  This Challenge generated a strong response from many research
groups, from both the computational geometry and the combinatorial
optimization communities, and resulted in a lively exchange of
solution ideas.

For CG Week 2020, the second CG:SHOP Challenge became an event within
the CG Week program, with top performing solutions reported in the
Symposium on Computational Geometry proceedings. The schedule for the
Challenge was advanced earlier, to give an opportunity for more
participation, particularly among students, e.g., as part of course
projects. 

The third edition of the Challenge in 2021 continued
the 2020 format, leading to contributions in the SoCG proceedings.

\section{The Challenge: Coordinated Motion Planning}

Coordinating the motion of a set of objects is a fundamental problem
that occurs in a large spectrum of theoretical contexts and practical applications.
A typical task arises from relocating a large collection of agents
from a given start into a desired goal configuration in an efficient manner,
while avoiding collisions between objects or with obstacles.

\subsection{The Problem}
\begin{figure}[h]
                \centering
                %\hfil
                        \includegraphics[width=.33\columnwidth]{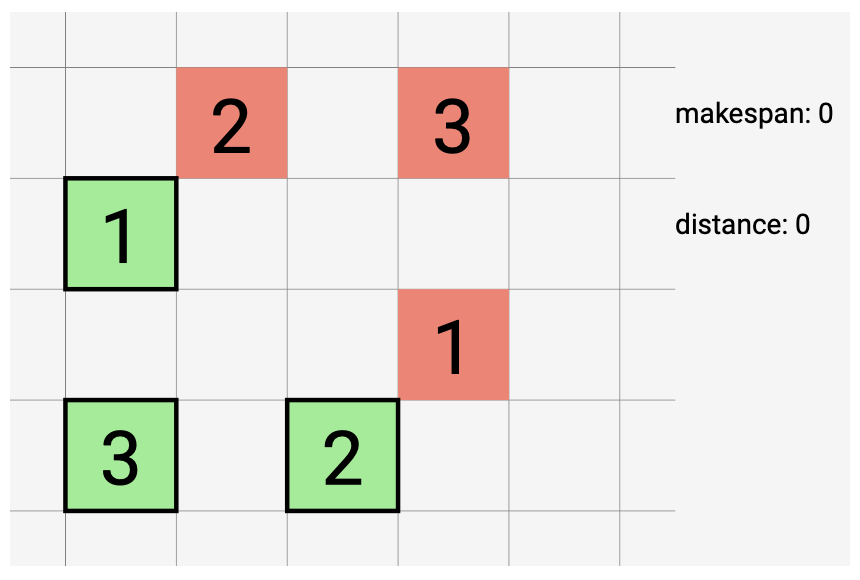}
                \hfil
                        \includegraphics[width=.33\columnwidth]{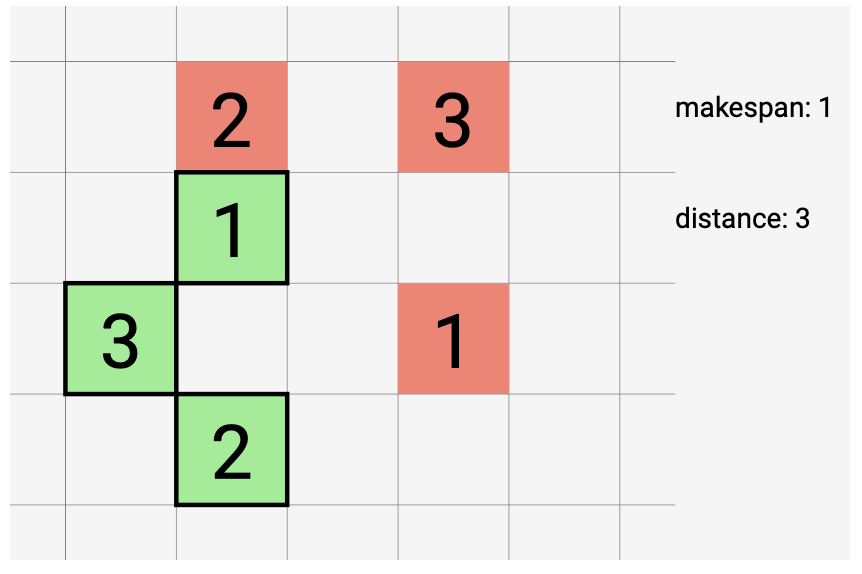}
                \hfil
                        \includegraphics[width=.33\columnwidth]{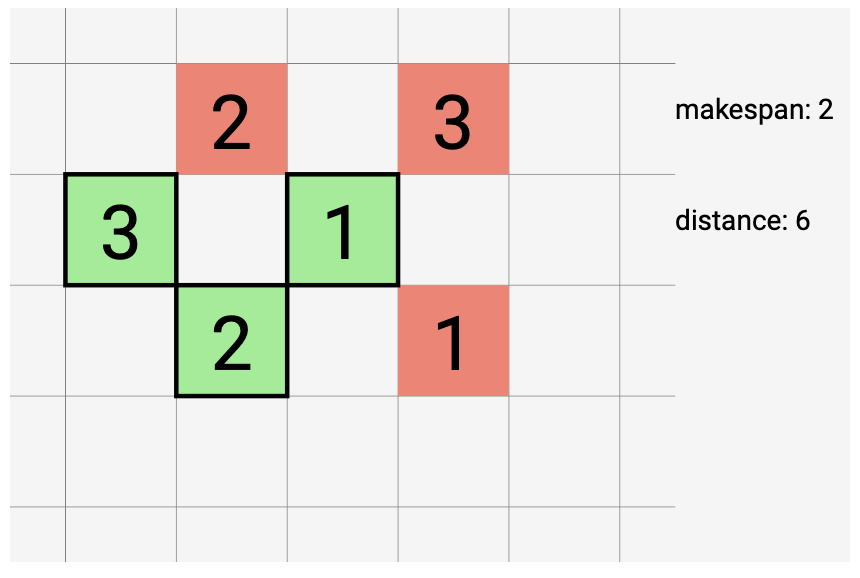}
                \hfil
                        \includegraphics[width=.33\columnwidth]{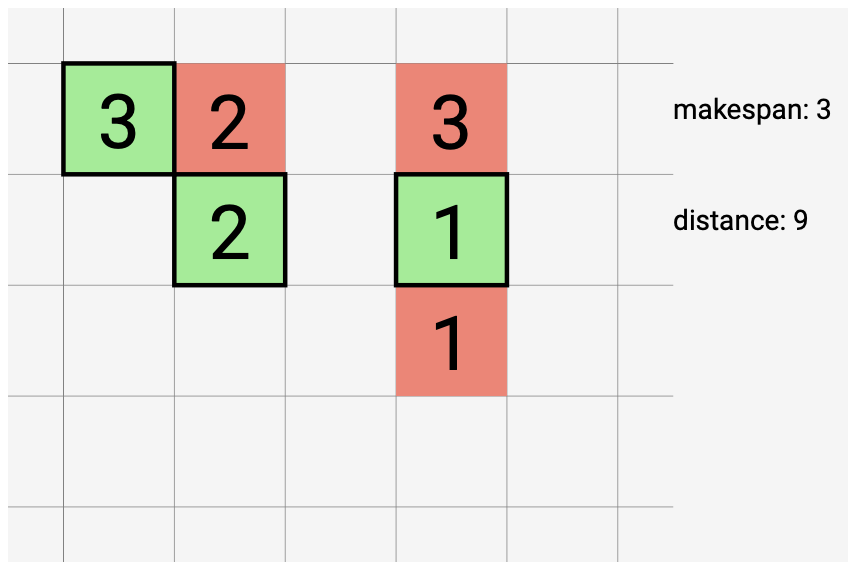}
                \hfil
                        \includegraphics[width=.33\columnwidth]{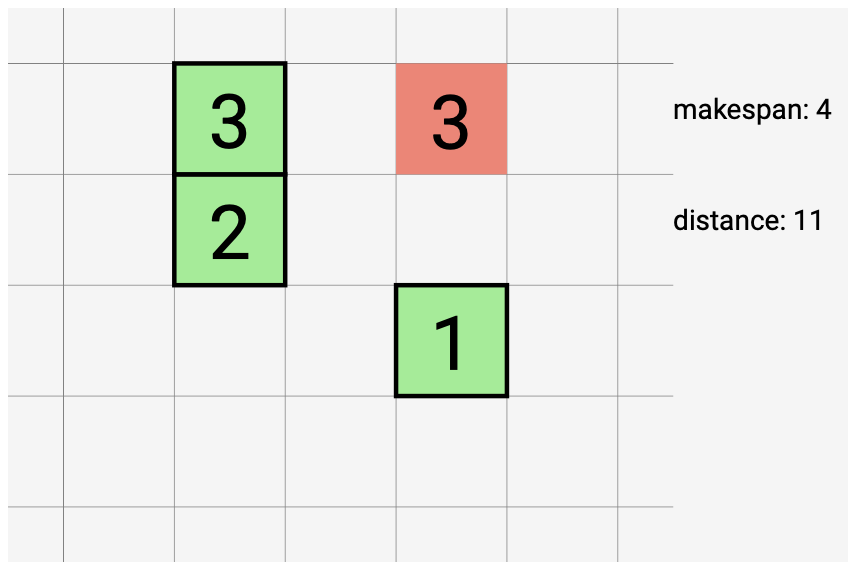}
                \hfil
                        \includegraphics[width=.33\columnwidth]{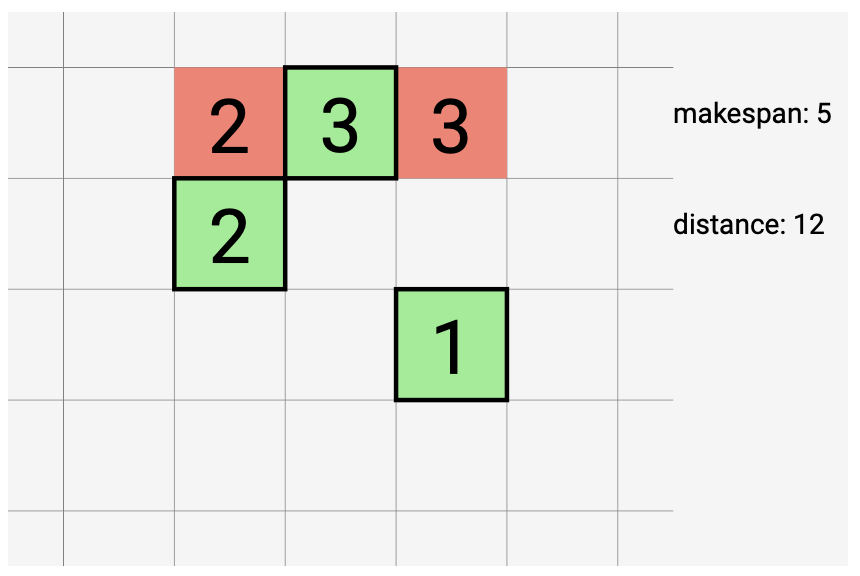}
                \hfil
                        \includegraphics[width=.33\columnwidth]{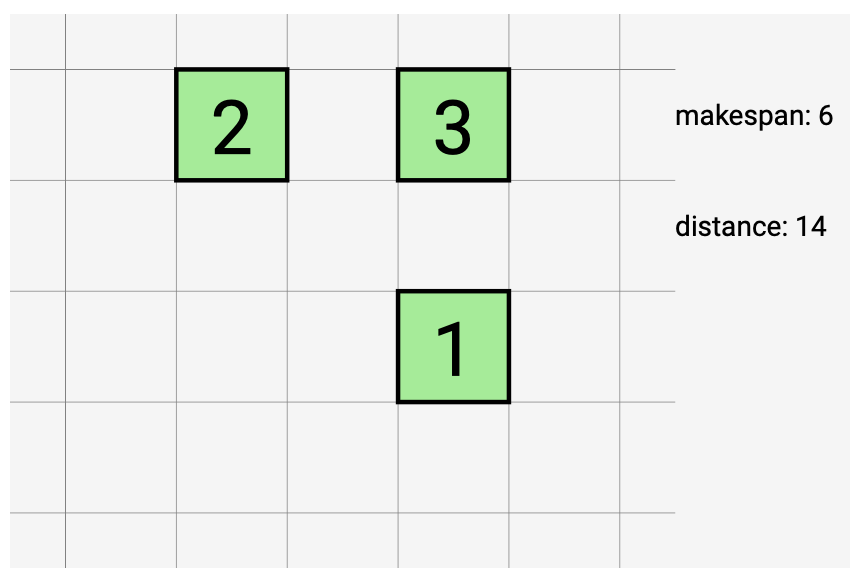}
                \hfil
                \caption{A feasible sequence of positions that minimizes both makespan and total distance. 
Robot positions are shown in green, target positions in red.}
                \label{fig:example}
\end{figure}

The specific problem that formed the basis of the 2021 CG Challenge was the following;
see Figure~\ref{fig:example} for a simple example.

\medskip
\textbf{Problem:} {\sc Coordinated Motion Planning} of unit squares in the plane.

\textbf{Given:} A set of $n$ axis-aligned unit-square robots in the plane, a set $S=\{s_1,\ldots,s_n\}$
of $n$ distinct \emph{start} pixels (unit squares) of the integer grid, and a
set $T=\{t_1,\ldots,t_n\}$ of $n$ distinct \emph{target} pixels of the integer
grid. In addition, there may be a set of obstacles,
consisting of a number of stationary, blocked pixels that cannot be used by
robots at any time.

\textbf{Goal:} The task is to compute a \emph{feasible} set of trajectories for all
$n$ robots, with the trajectory for robot $i$ moving it from $s_i$ to $t_i$, such that
the overall schedule is optimal with respect to an objective function.

\medskip
In order to be feasible, a trajectory must satisfy a number of conditions.
During each unit of time, each robot can move (at unit speed) in a direction 
(north, south, east or west) to an adjacent pixel, provided the robot remains 
disjoint from all other robots during the motions.
This condition has to be satisfied at all times, not just when robots are at
pixel positions. For example, if there are robots at each of the two adjacent
pixels $(x,y)$ and $(x + 1,y)$, then the robot at $(x,y)$ can move east
into position $(x + 1,y)$ only if the robot at $(x + 1,y)$
moves east at the same time, so that the two robots remain in contact, during the movement, but never overlap.

\medskip
The contest was run on two different objective functions, as follows.

\begin{description}
	\item[MAX:] Minimize the makespan, i.e., the time until all robots have reached their destinations.
	\item[SUM:] Minimize the total distance traveled by all robots.
\end{description}

\subsection{Related work}
Coordinating the motion of many agents plays a central role when
dealing with large numbers of moving robots, vehicles, aircraft, or people.
How can each agent choose an efficient
route that avoids collisions with other agents
as they simultaneously move to their destinations?
These basic questions arise in many applications,
such as ground swarm robotics~\cite{rubenstein2014programmable,sw-sr-08},
aerial swarm robotics~\cite{cpdsk-saesr-18,kumar},
air traffic control~\cite{delahaye2014mathematical}, and
vehicular traffic networks~\cite{fhtwhfe-mift-11,ss-hbtn-04}.

{Multi-robot coordination} dates back to the
early days of robotics and computational geometry.
The seminal
work by Schwartz and Sharir~\cite{ss-pmpcbpb-83} from the 1980s
considers coordinating the motion of disk-shaped objects
among obstacles.  Their algorithms are polynomial in the
complexity of the obstacles, but exponential in the number of disks.
Hopcroft et al.~\cite{hss-cmpmio-84} and Hopcroft and
Wilfong~\cite{hw-rmompgs-86} proved \PSPACE-completeness of
moving multiple robots to a target configuration,
showing the significant challenge of coordinating
many robots.

There is a vast body of other related work dealing with multi-robot motion planning, both from
theory and practice. %a practical and a theoretical angle, including discrete and continuous geometric variants.
For a more extensive overview, see~\cite{dfk+-arxiv}.
%Here we only mention a particular aspect that distinguishes this paper from~\cite{dfk+-arxiv}, as follows.
In both discrete and geometric variants of the problem, the objects
can be \emph{labeled}, \emph{colored}, or \emph{unlabeled}.
In the \emph{labeled} case, the objects are all distinguishable and each object has its own, uniquely defined target position.
%This is the most extensively studied scenario among the three.
In the \emph{colored} case, the objects are partitioned into $k$ groups and each target position can only be covered by an object with the right color.
This was considered by Solovey and Halperin~\cite{sh-kcmrmp-14}, who present and evaluate a practical sampling-based algorithm.
In the \emph{unlabeled} case, objects are indistinguishable and target positions can be covered by any object.

This scenario was first considered by Kloder and Hutchinson~\cite{kh-ppimf-06}, who presented a practical sampling-based algorithm.
Turpin~et~al.~\cite{tmk-tpams-13} give an algorithm for finding a solution in polynomial time, if one exists.
This is optimal with respect to the longest distance traveled by any one robot,
but only holds for disk-shaped robots under additional restrictive assumptions on the free space.
For unit disks and simple polygons, Adler et al.~\cite{adh+-emmpudsp-15} provide a {polynomial-time} algorithm under the additional assumption that the start and target positions have some minimal distance from each other.
Under similar separability assumptions, Solovey et al.~\cite{syz+-mpudog-15} provide a polynomial-time algorithm that produces a set of paths that is no longer than $\mbox{OPT}+4m$, where $m$ is the number of robots and $\mbox{OPT}$ is the total length of the paths in an optimal solution.
However, they do not consider the makespan, but only the total path length.
On the negative side, Solovey and Halperin~\cite{sh-hummp-15} prove that the unlabeled multiple-object motion planning problem is \PSPACE-hard, even when restricted to unit square objects in a polygonal environment.

There is also a wide range of practical related work.
Self-configuration of robots as active agents was studied by Naz et al.~\cite{naz2016distributed}.
A basic model in which robots are used as building material was introduced by Derakhshandeh et
al.~\cite{derakhshandeh2015algorithmic,derakhshandeh2016universal}.
This resembles Claytronics robots like Catoms; see Goldstein and Mowry~\cite{goldstein2004claytronics}.
In more recent work, Thalamy et al.~\cite{thalamy2019distributed} consider using scaffolding structures for asynchronous
reconfiguration.

%Most other previous work on coordinated motion planning has largely focused on
%{sequential} schedules, where one robot moves at a time, with objectives
%such as minimizing the number of moves.  In practice, however, robots usually
%move simultaneously, so we desire a \emph{parallel motion schedule},
%with the objective of minimizing the completion time,
%called \emph{makespan}. How well can we exploit parallelism in
%a robot swarm to achieve an efficient schedule?

For an instance of parallel reconfiguration, a lower bound
for the time required for \emph{all} robots to reach their destinations is the
time it takes to move just \emph{one} robot to its
destination in the absence of other robots,
i.e., by the maximum distance between a robot's origin and destination.
Moving a dense arrangement of robots to their destinations while avoiding
collisions may require substantially more time than this lower bound.
This motivates the \emph{stretch factor}, which is defined to be the ratio of the time taken by
a parallel motion plan divided by the simple lower bound.

\begin{figure}[h]
                \centering
                \hfil
                        \includegraphics[width=.16\columnwidth]{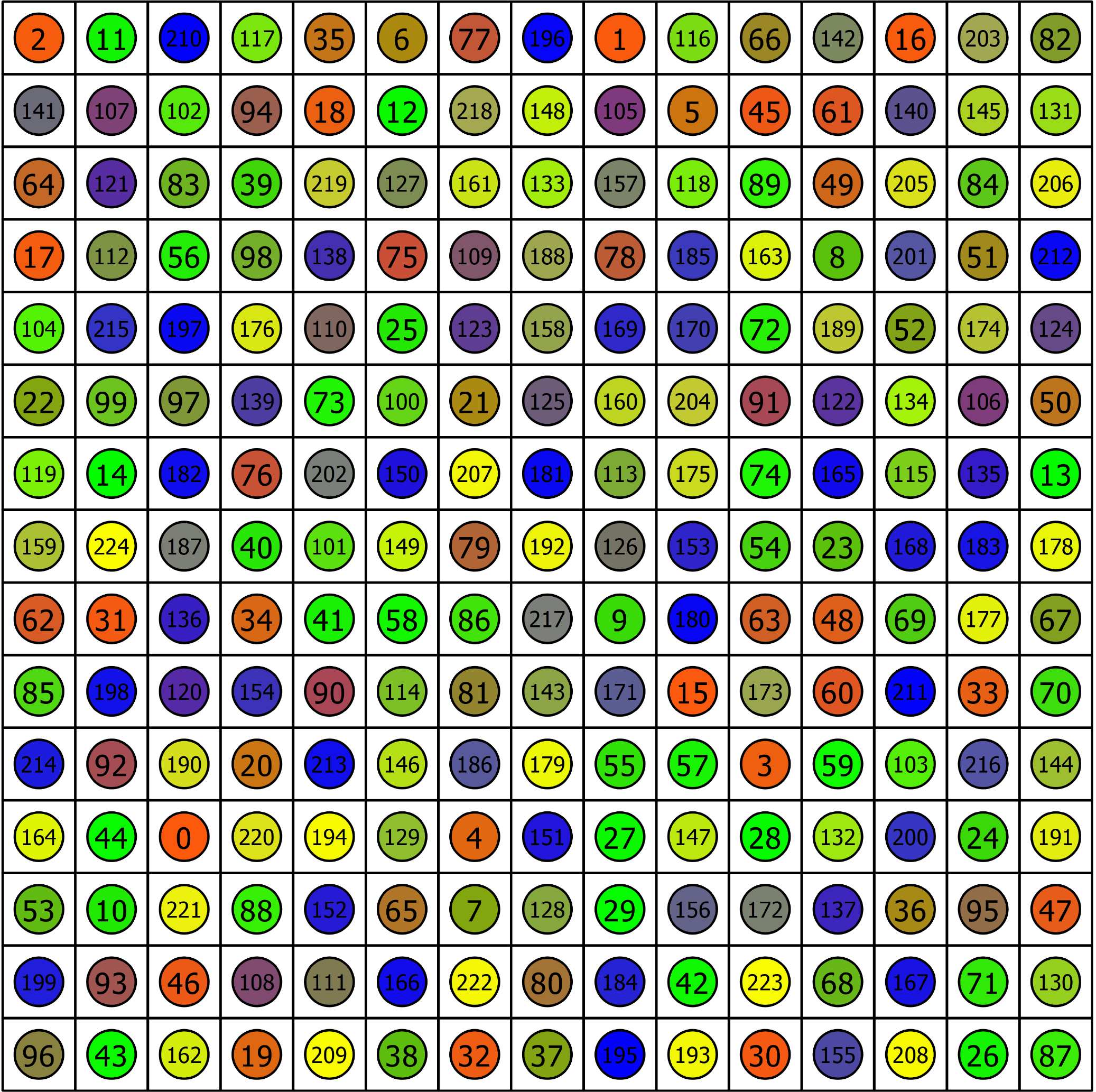}
                \hfil
                        \includegraphics[width=.16\columnwidth]{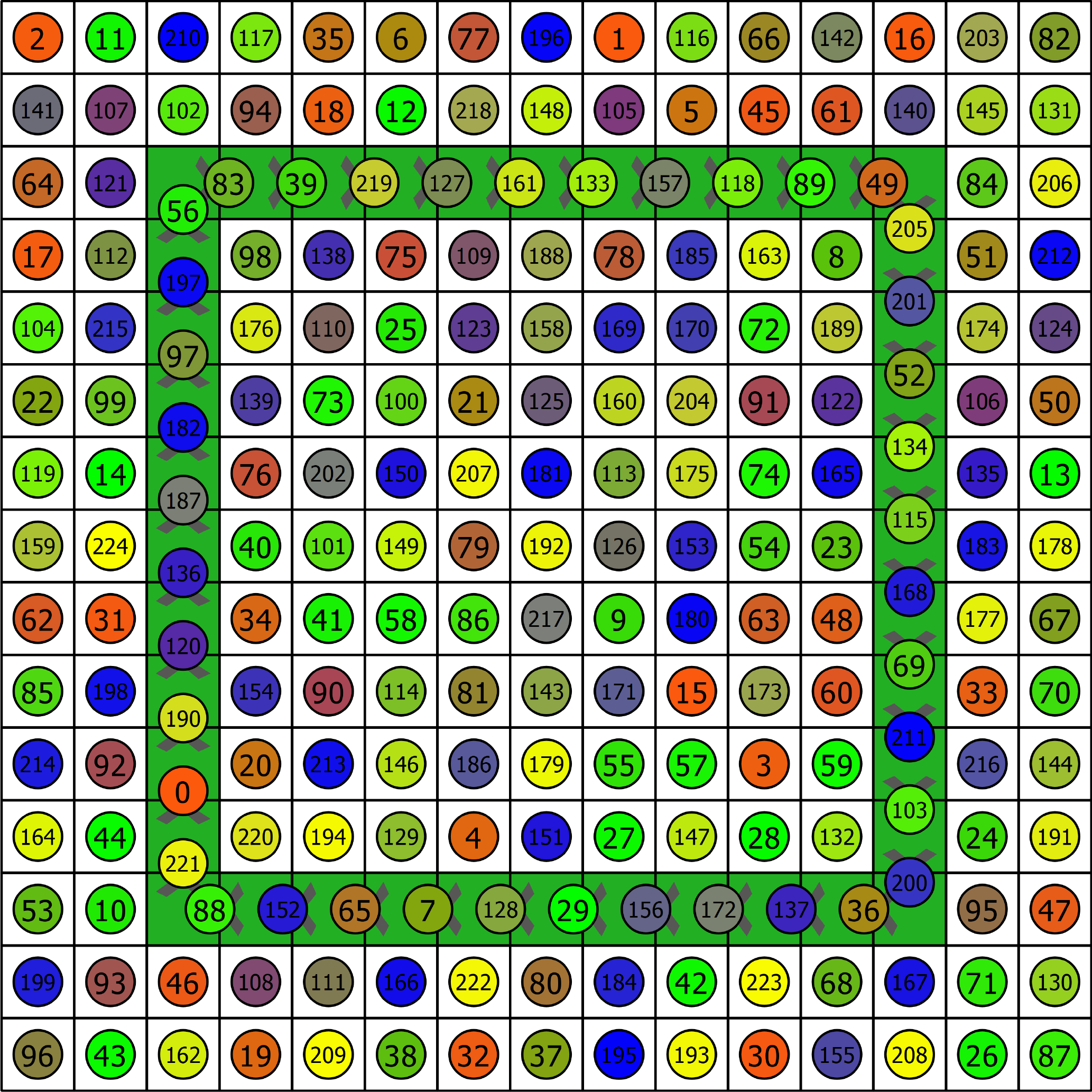}
                \hfil
                        \includegraphics[width=.168\columnwidth]{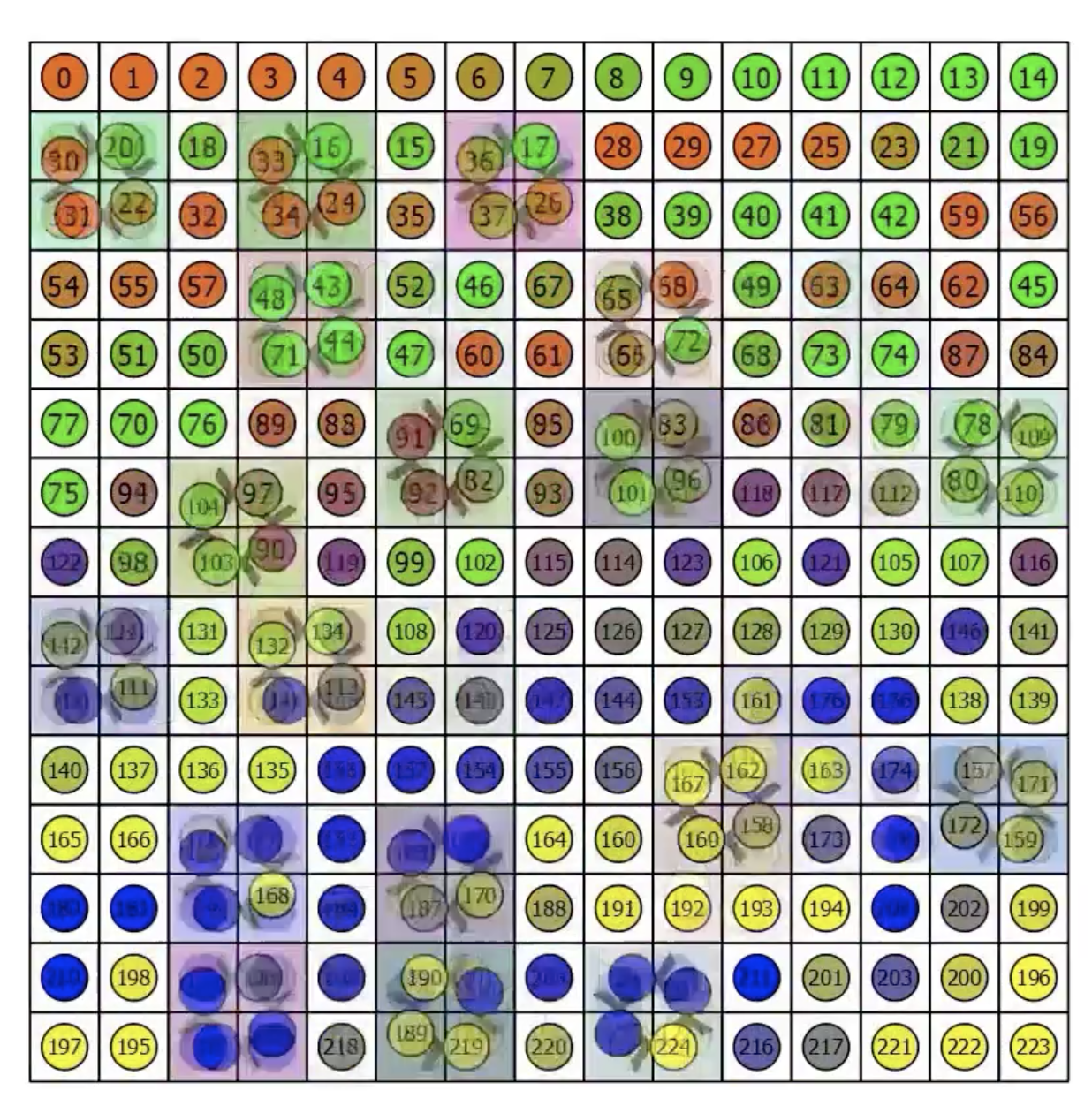}
                \hfil
                        \includegraphics[width=.16\columnwidth]{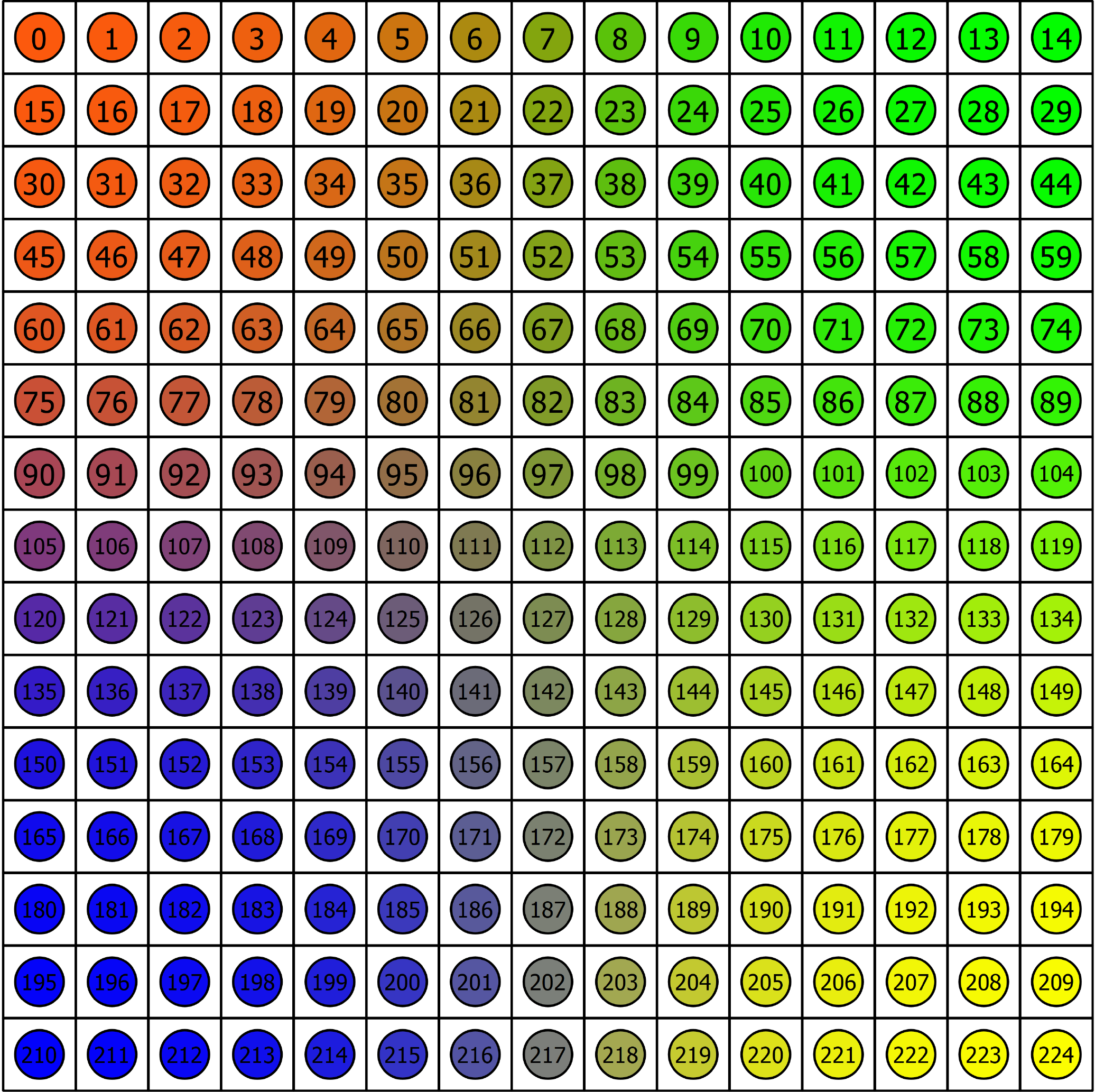}
                \hfil
                \caption{Parallel reconfiguration, established by~\cite{dfk+-cmprs-18,dfk+-arxiv}:
(a) Start configuration. (b) A feasible, parallel reconfiguration move. (c)
Parallel reconfiguration moves. (d)
Target configuration.
See \protect\url{https://www.ibr.cs.tu-bs.de/users/fekete/Videos/CoordinatedMotionPlanning.mp4}
for a video~\cite{coordinated_video}%with more detailed animations
.}
                \label{fig:coordinated}
\end{figure}

In recent work, Demaine et al.~\cite{dfk+-cmprs-18,dfk+-arxiv}
provide several fundamental insights into
these problems of coordinated motion planning for the scenario with labeled robots.
They develop algorithms that
(under relatively mild assumptions on the separation between robots, slightly more generous than
provided in the Challenge) achieve
\emph{constant} stretch factors that are independent of the
number of robots. Thus, these algorithms provide an absolute performance guarantee on the
makespan of the parallel motion schedule, which implies that the schedule is
a constant-factor approximation of the best possible schedule.
For densely packed arrangements of robots (without separation assumptions),
they proved that a constant stretch factor is no longer possible, and gave
upper and lower bounds on the worst-case stretch factor. See Figure~\ref{fig:coordinated}
for an illustration.

\subsection{Instances}
An important part of any challenge is the creation of suitable instances.
If the instances are easy to solve to optimality, the challenge becomes trivial.
If instances require a huge amount of computation to find any decent solutions,
the challenge may heavily favor teams that can afford better computation equipment.
The same is true if the set of instances becomes too large to manage with a single (or few) computers.

A priori, it is difficult to tell how hard finding a good solution to an instance is and which parameters
influence the difficulty of solving an instance (and in what way).
Therefore, it is important to create a set of instances that are diverse with respect to parameters that are
likely to influence their difficulty.

To create interesting and challenging instances, we thus developed an instance generator that can be tuned by 
adjusting several parameters to create diverse instances.
Basic parameters control the size of the map, the density to which the map is filled with robots,
and the distribution functions for start and target positions.
For the distribution functions, we used uniform distributions and distributions based on various images, such as microscope images of bacteria colonies.
The image-based instances often have interesting, natural patterns; their accumulations of robots in some areas were expected to yield challenging instances.
Additionally, we created artificial bottlenecks and accumulations of robots by inserting obstacles and clusters of robots to increase the difficulty.

\begin{figure}[ht]
	\centering
	\includegraphics[width=.49\textwidth,clip,trim=2.5cm 0cm 2.5cm 0cm]{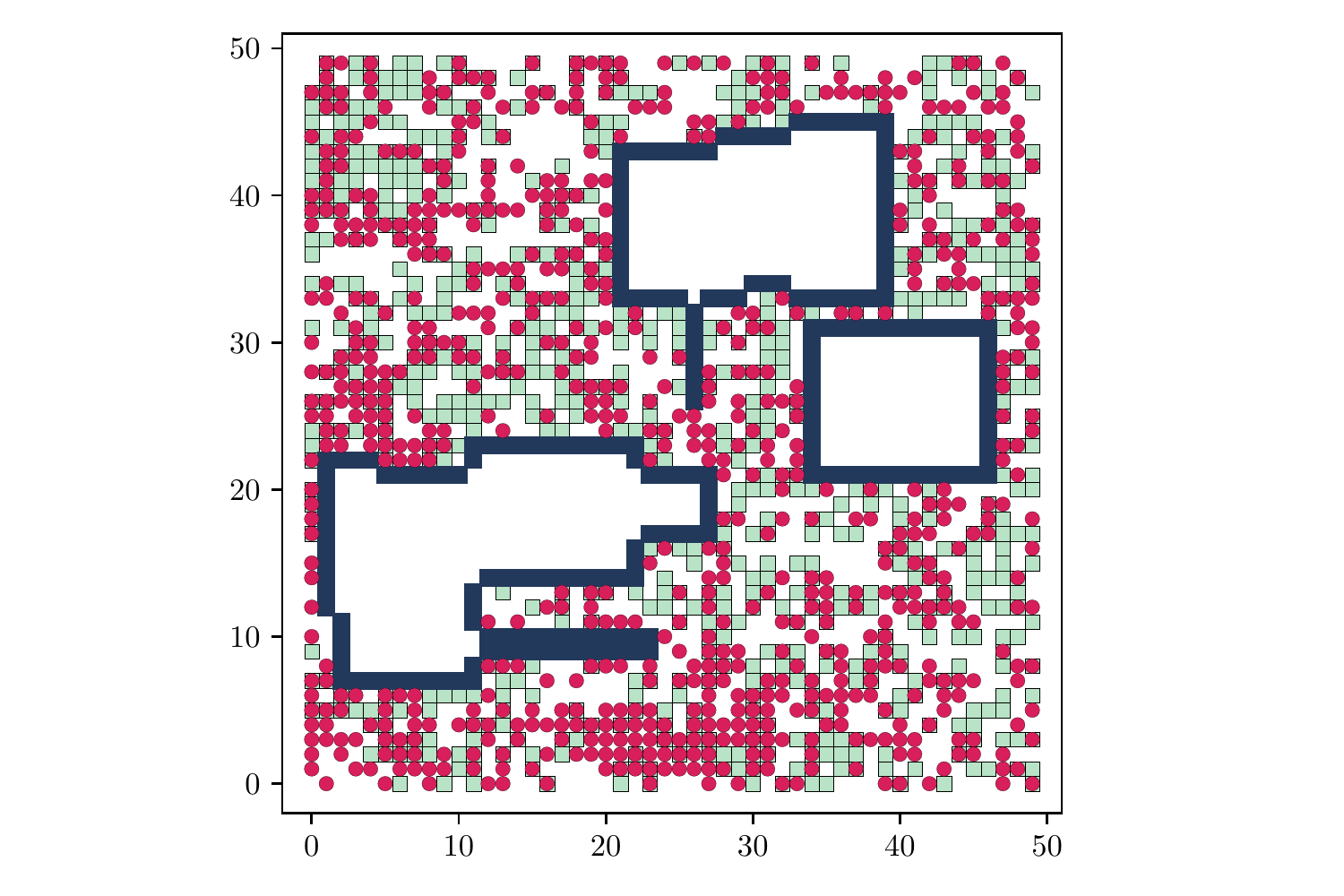}\hfil
	\includegraphics[width=.49\textwidth,clip,trim=2.5cm 0cm 2.5cm 0cm]{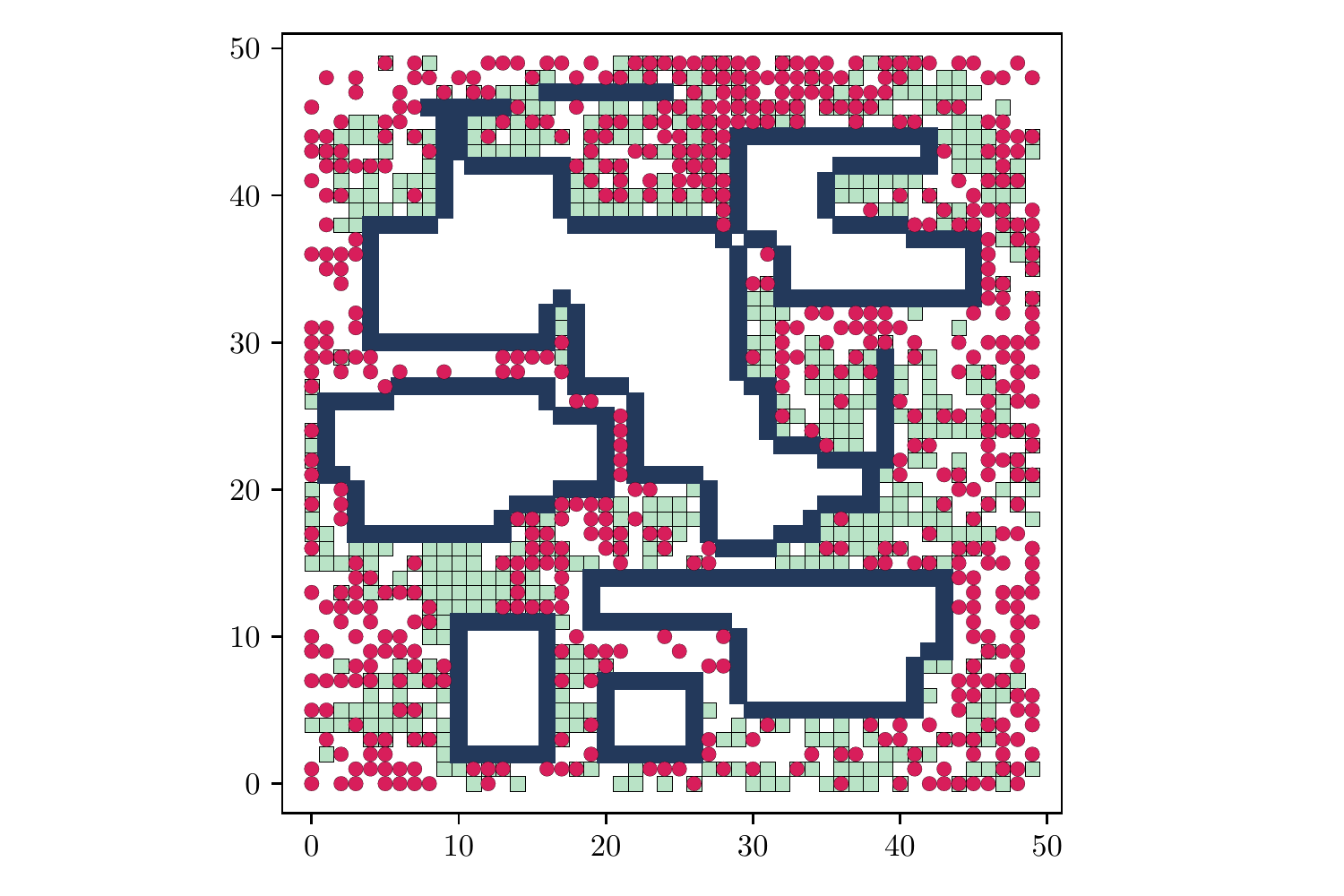}
	\caption{A visualization of two instances with obstacles (dark pixels).
	         Start and target positions of robots are indicated by red disks and green squares.}
	\label{fig:obstacle-example}
\end{figure}

Obstacles are created by randomly placing rectangles with (truncated) normally distributed sizes.
Any area that is completely surrounded by obstacles also becomes an obstacle to ensure that instances remain feasible.
Despite its simplicity, this procedure is able to create various interesting obstacles after some parameter tuning; see Figure~\ref{fig:obstacle-example} for an example.

The idea of robot clusters is to have groups of robots
that are close to each other in the start and target configurations,
but have to travel a significant distance as a group.
During its motion, such a robot cluster may have to pass through narrow obstacles
or may clash with other clusters, yielding interesting conflicts.
Robot clusters are created by specifying the desired number of clusters
and the parameters of a normal distribution controlling their size.
For each cluster, a start and a target is randomly selected according to
the distribution functions controlling the robots' start and target positions.
A corresponding number of robots is then distributed within
size-dependent windows around the start and target positions of the cluster.
These windows may be too small to contain the start and target positions due to obstacles or 
robots that have already been placed; in that case, we retry with a larger window.

As described above, certain parameters (such as the density) can be controlled directly by our generator.
To ensure that our instances are diverse even with respect to parameters for which this is not possible, we generated
a large set of instances by selecting a diverse set of possible values for each parameter of our generator
and then generating several instances for each possible combination of these parameter values.

\begin{figure}[ht]
	\centering
	\includegraphics[width=.99\textwidth]{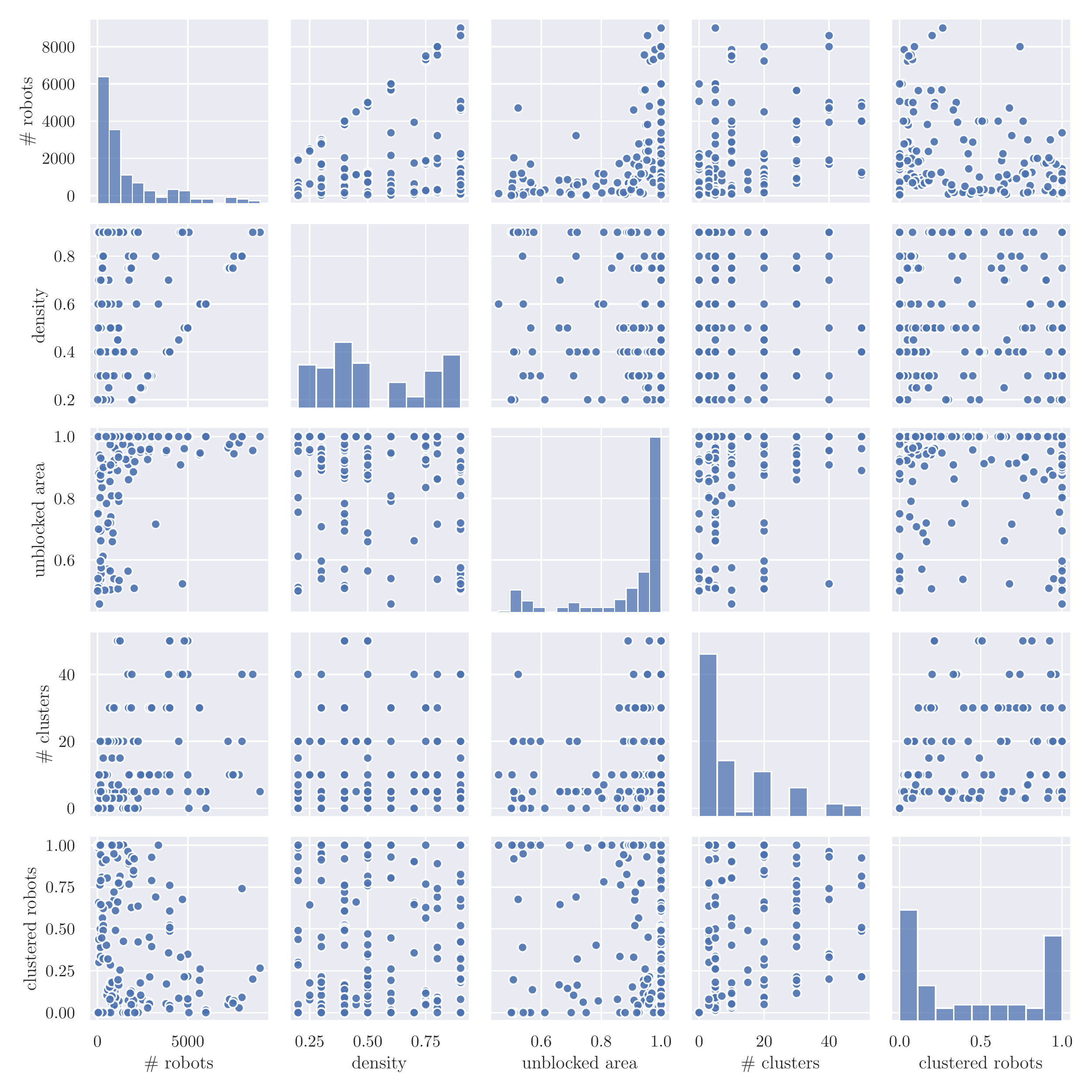}
	\caption{A pair plot of the distribution of our instances according to several important instance properties.}
	\label{fig:instance-distribution}
\end{figure}

This resulted in a set of more than 10000 candidate instances from which we then selected the challenge instances as follows.
We measured several properties for each of the generated instances, including the number of robots, the density, the number of robot clusters, the number of robots that were part of a robot cluster and the volume and free area of the map.
Based on these properties, we then defined and tuned a distance function between the instances and used a
greedy dispersion algorithm to select a total of 200 diverse instances from the candidates;
see Figure~\ref{fig:instance-distribution} for an overview of the distribution of some important
properties across the selected instances.

\subsection{Evaluation}
%<<<<<<< HEAD
%The contest was run on a total of 203 instances. For either objective function,
%the achieved score for any instance was a number between 0 and 1, with 1
%corresponding to an optimal score and 0 being a default score; as a
%consequence, the total score became a number between 0 and 203 for either contest,
%with 203 being best possible. For each contest (MAX and SUM), participants were
%compared based on their total scores; the winner for each contest was the one
%with the highest score.

%For each instance and either objective, the score was defined as the ratio $\frac{L}{V}$ 
%between a lower bound $L$ on the objective function and the achieved 
%objective value $V$ for that instance. During the course of the contest
%(before final results were known), 
%a useful valid lower bound for MAX was the maximum Manhattan distance
%between start and target position for any robots in an instance. For SUM, a
%similarly useful valid lower bound was the sum of all Manhattan distances between start and
%target position for any robots in an instance. Thus, the inverse of the score
%corresponded to the (maximum or average) stretch factor for motion planning; the
%inverse was used for normalization and comparability between instances.
%
%For final scoring (and deciding the contest), the best feasible value $V^*$
%achieved by any contestant was used as the value $L$. Thus, the best solution for any
%instance received a score of 1. This ensured that instances with high stretch
%factor get similar weight, so potentially difficult instances receive scores of
%similar size.
%=======
The contest was run on a total of 203 instances, 3 of which were small, manually generated instances.
For either objective function, a team's score for an instance $I$ is
the ratio $\nicefrac{L}{V}$ between $L$, the objective value of the best submitted solution for $I$,
and $V$, the best objective value of any valid solution for $I$ submitted by the team.
The score for each instance is thus a number
between 0 and 1, with 1 being the score awarded to the teams with the best
submitted solution and 0 being a default score.
As a consequence, the total score is a number between 0 and 203 for
either contest, where 203 is best possible. For each contest (MAX and SUM),
participants were compared based on their total scores; the winner for each
contest was the one with the highest score.
%>>>>>>> dc0be36 (Some fixes.)

In case of ties, the tiebreaker was set to be the time a specific score was obtained. 
This turned out not to be necessary.

\subsection{Categories}
The contest was run in an \emph{Open Class}, in which participants could use any
computing device, any amount of computing time (within the duration of the
contest) and any team composition. In the \emph{Junior Class}, a
 team was required to consist exclusively of participants who were eligible
according to the rules of CG:YRF (the \emph{Young Researchers Forum} of CG Week), 
defined as not having defended a formal doctorate before 2019.

\subsection{Server and Timeline}
The contest itself was run through a dedicated server at TU Braunschweig,
hosted at \url{https://cgshop.ibr.cs.tu-bs.de/competition/cg-shop-2021/}.
It opened at  00:00 AoE on November 20, 2020, and closed
at 24:00 (midnight, AoE), February 15, 2021. 

\section{Outcomes}
A total of 17 teams submitted solutions. In the end, the leaderboard
for the top 10 teams in both categories
looked as shown in Table~\ref{tab:top10_MAX} and Table~\ref{tab:top10_SUM}; 
note that according to the scoring function, a higher score is better.

\begin{table}[h!]
  \begin{center}
    \caption{The top of the final leaderboard for MAX. ``Best solutions'' are the best found by any participating team,
which does not exclude the possibility of better solutions, but provides a score of 1 for that instance; scores are truncated to four
decimal places. Junior teams are indicated by $^J$.}
    \label{tab:top10_MAX}
    \begin{tabular}{|r|l|r|r|} 
      \hline
      \textbf{Position} & \textbf{Team} & \textbf{Score} & \textbf{\# best }  \\
      			& 		& 		 & \textbf{solutions} \\
      \hline
      1 & Shadoks	& 202.9375	& 202 \\
      2 & UNIST		& 174.0180	& 14  \\
      3 & gitastrophe$^J$	& 159.5472	& 24 \\
      4 & S10ppy J035$^J$	& 109.2778	& 3 \\
      5 & \'Ecole Polytechnique$^J$ & 90.8213 & 3 \\
      6 & TUeSWarM$^J$	& 83.0897	& 7 \\
      7 & JoJo$^J$		& 75.2449	& 6 \\
      8 & BlueTeamTechnion$^J$ & 65.0906	& 7 \\
      9 & Lasteam$^J$	& 41.6065	& 1 \\
      10 & Kleinkariert$^J$	& 36.0898	& 0 \\
      \hline
    \end{tabular}
  \end{center}
\end{table}

\begin{table}[h!]
  \begin{center}
    \caption{The top of the final leaderboard for SUM. ``Best solutions'' are the best found by any participating team,
which does not exclude the possibility of better solutions, but provides a score of 1 for that instance; scores are truncated to four
decimal places. Junior teams are indicated by $^J$.}
    \label{tab:top10_SUM}
    \begin{tabular}{|r|l|r|r|} 
      \hline
      \textbf{Position} & \textbf{Team} & \textbf{Score} & \textbf{\# best }  \\
      			& 		& 		 & \textbf{solutions} \\
      \hline
      1 & gitastrophe$^J$	& 198.4943	& 57 \\
      2 & UNIST			& 191.7893	& 120 \\
      3 & Shadoks		& 180.4952	& 0 \\
      4 & cgi@tau		& 175.0754	& 6 \\
      5 & BlueTeamTechnion$^J$ 	& 165.5633	& 34 \\
      6 & TUeSWarM$^J$		& 146.2244	& 1 \\
      7 & S10ppy J035$^J$	& 133.450	& 1 \\
      8 & Team ITI$^J$		& 109.4631	& 0 \\
      9 & Kleinkariert$^J$	& 81.7086	& 0 \\
      10 & \'Ecole Polytechnique$^J$ & 75.4734	& 0 \\
      \hline
    \end{tabular}
  \end{center}
\end{table}

\begin{figure}
  \centering
  \includegraphics[width=0.9\textwidth]{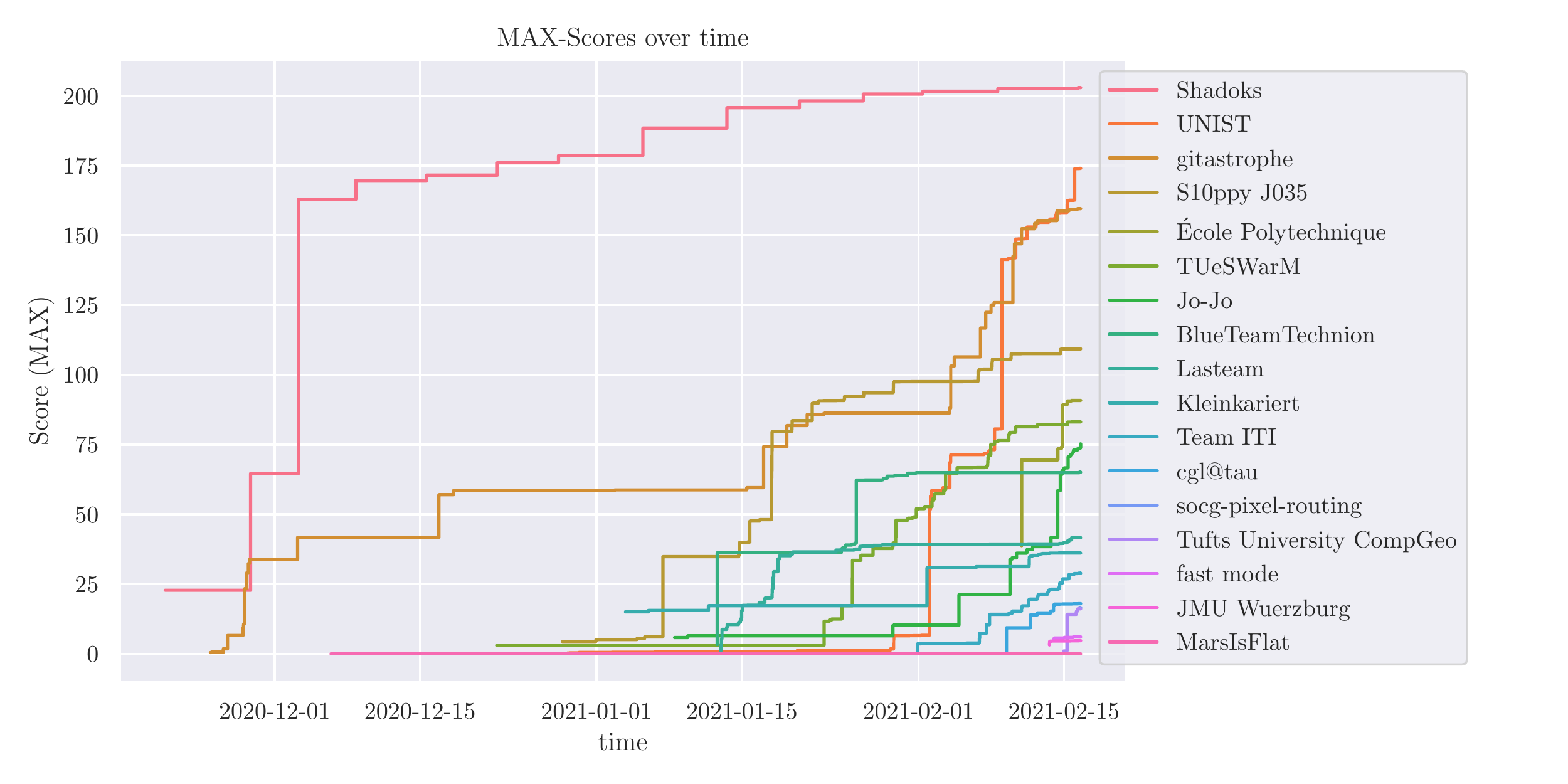}
  \includegraphics[width=0.9\textwidth]{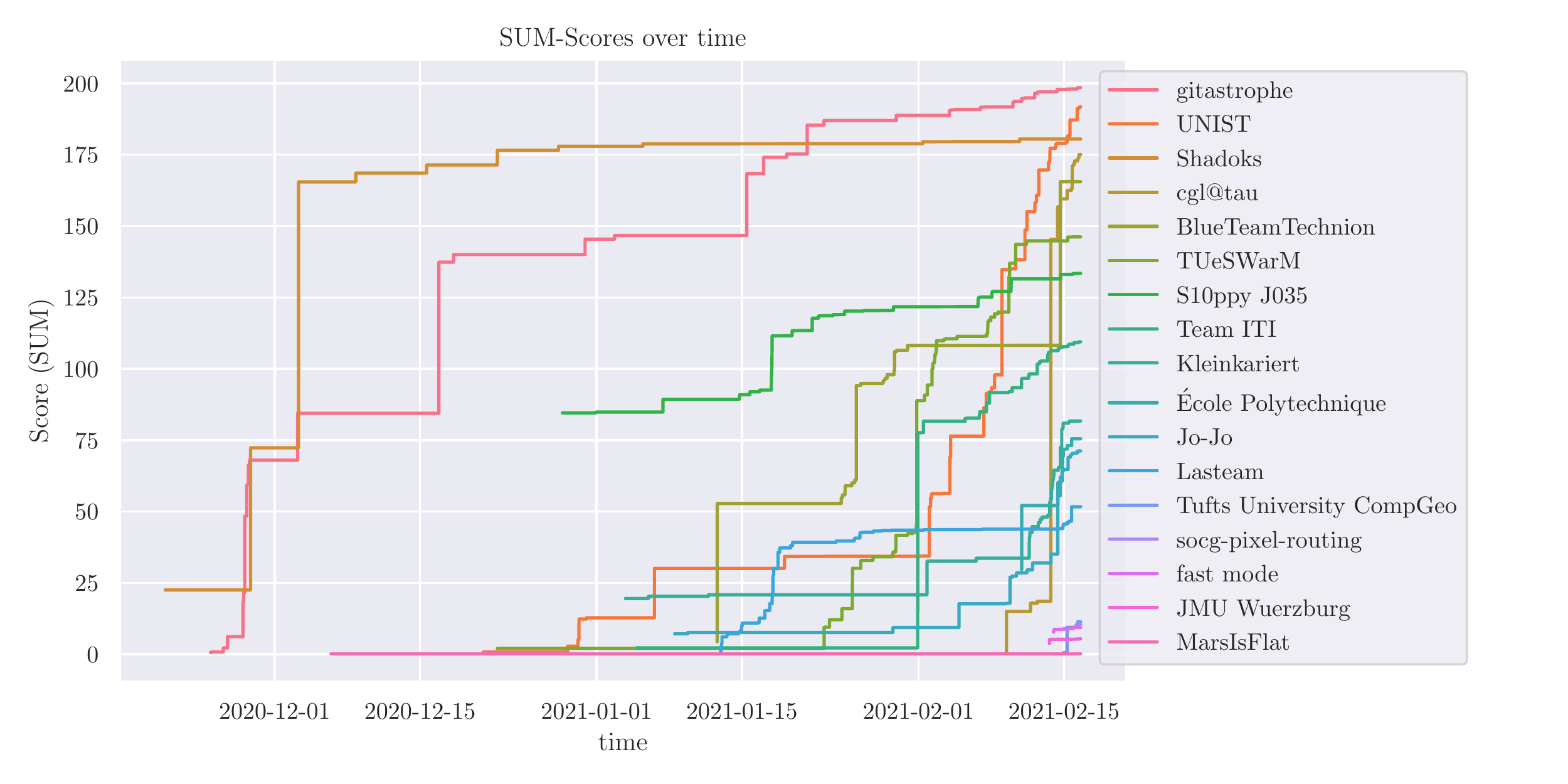}
  \caption{Progress of the teams over time. Team Shadoks built its lead in the
MAX-objective very early and only did minor improvements over the rest of the
time. For team gitastrophe, a continuous progress is visible. Team UNIST, on
the other hand, started relatively late. Most teams seem to have only focused
on the SUM-objective, as these scores show the largest jumps, with submissions
picking up in the last month of the competition.}
\label{fig:top10}
\end{figure}

\begin{figure}[]
  \centering
  \includegraphics[height=5cm]{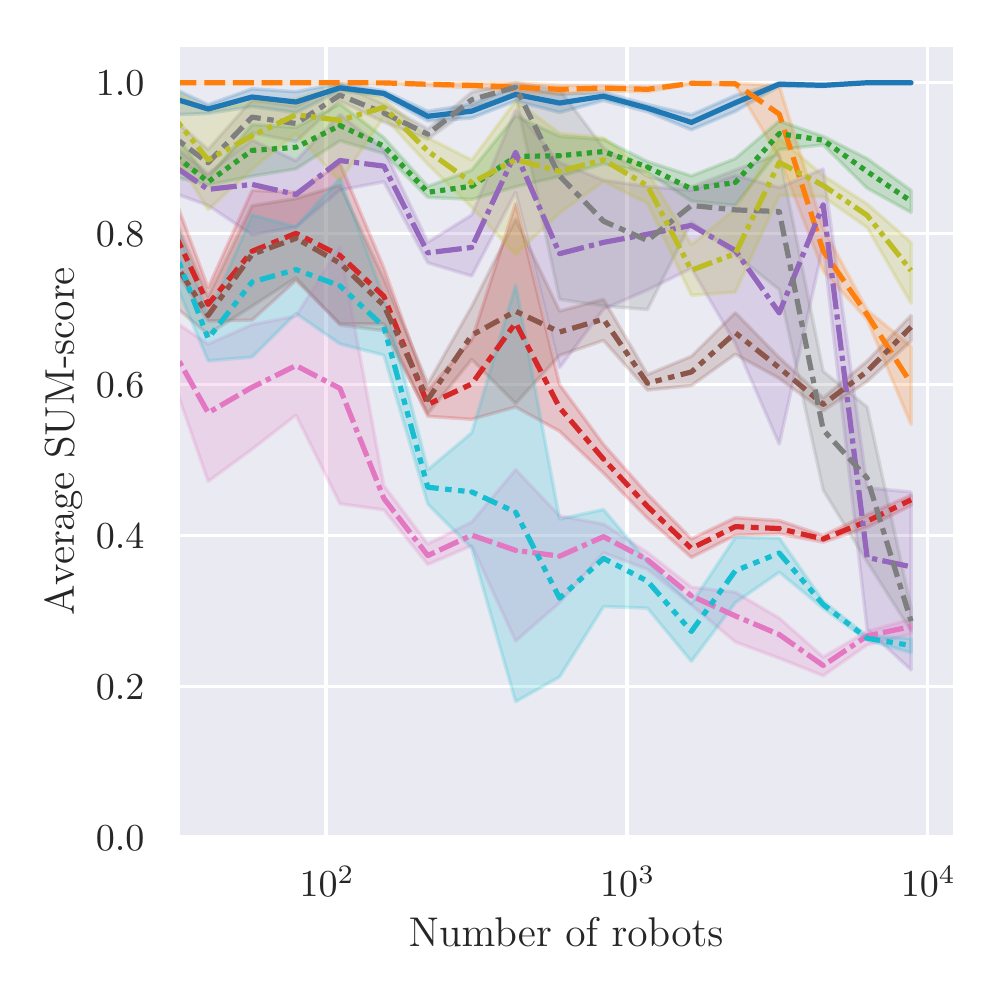}
  \includegraphics[height=5cm]{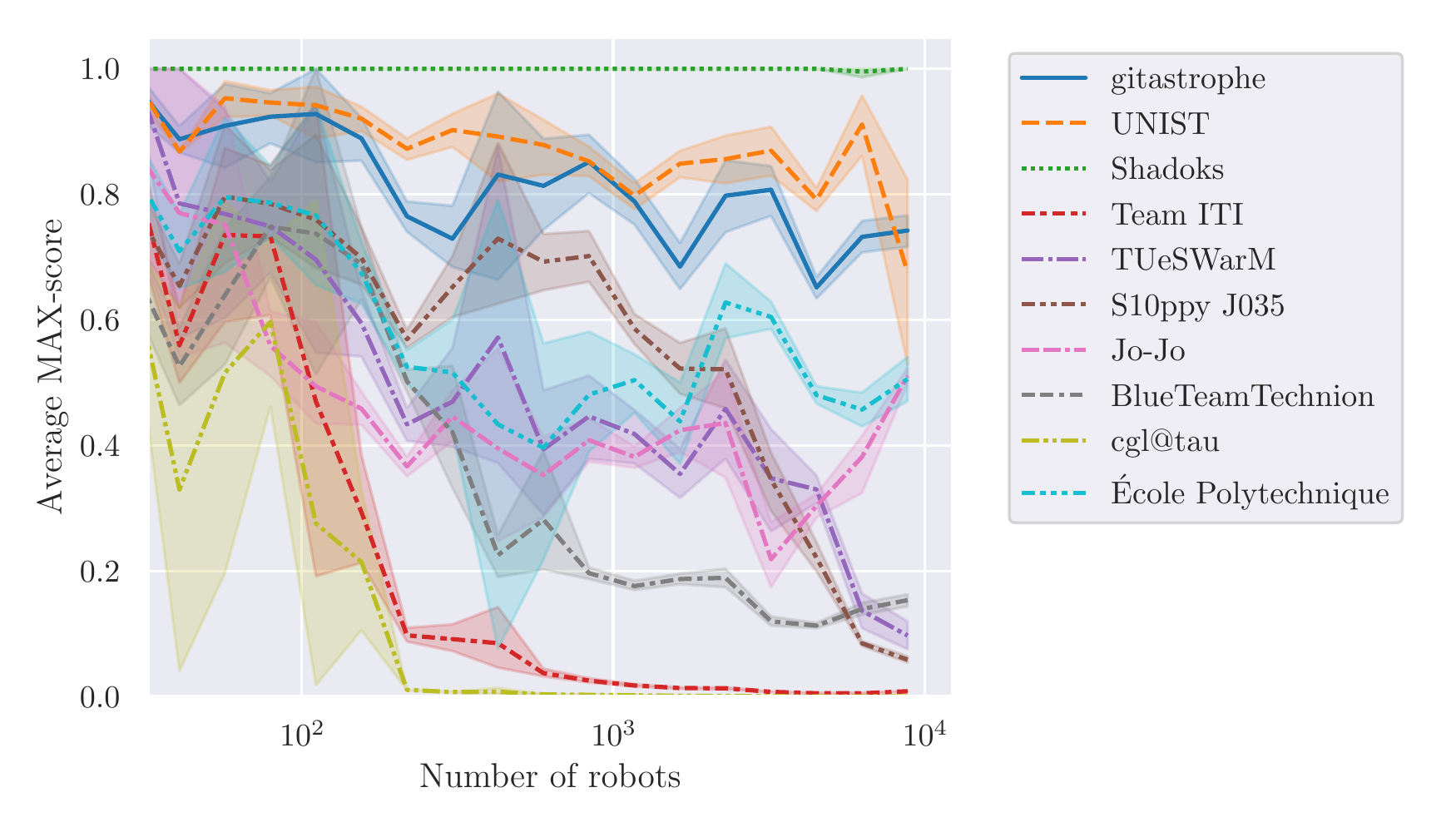}
  \caption{The scores over the instance sizes of the ten best teams. We can see
that team Shadoks is nearly unbeaten on the MAX objective. Team UNIST is very
strong for smaller instances and SUM, but the approach of team gitastrophe seems to
scale better and wins on the larger instances by a significant margin.}
\label{fig:scores_over_size}
\end{figure}

The progress over time of each team's score can be seen in Figure~\ref{fig:top10};
the best solutions for all instances (displayed by score) can be seen in Figure~\ref{fig:scores_over_size}.

There were three clear frontrunners, with Team Shadoks placing first and third in MAX and SUM,
respectively; Team UNIST came in second in both categories, while Team gitastrophe
placed third and first in MAX and SUM. A closer look at the scores reveals that 
Team Shadoks achieved almost perfect overall in MAX, which was sufficient
to place first in the sum of both objective functions, at 401.4318. 
Team UNIST managed a balanced outcome for both objective functions,
for a combined score of 365.8073. Conversely, Team gitastrophe achieved
the best score for SUM, but a slightly inferior result for MAX, resulting
in a combined score of 358.0415; at the same time, they also placed first
in all three events in the Junior Class.

These top 3 finishers were invited for contributions in the 
2020 SoCG proceedings, as follows.

\begin{enumerate}
\item Team Shadoks: 	Lo\"ic Crombez, Guilherme D.~da Fonseca, Yan Gerard, Aldo Gonzalez-Lorenzo, Pascal Lafourcade, Luc Libralesso~\cite{shadoks}.
\item Team UNIST: 	Hyeyun Yang, Antoine Vigneron~\cite{UNIST}.
\item Team gitastrophe: Jack Spalding-Jamieson, Paul Liu, Brandon Zhang, Da Wei Zheng~\cite{gitastrophe}.
\end{enumerate}

All three teams engineered their solutions based on a spectrum of heuristics for generating start configurations,
in combination of a variety of local search methods, including simulated annealing, $k$-optimization, 
and more refined, tailor-made approaches.
Details of their methods and the engineering decisions they made are given in their respective papers.

\begin{figure}[]
  \centering
  \includegraphics[width=0.95\textwidth]{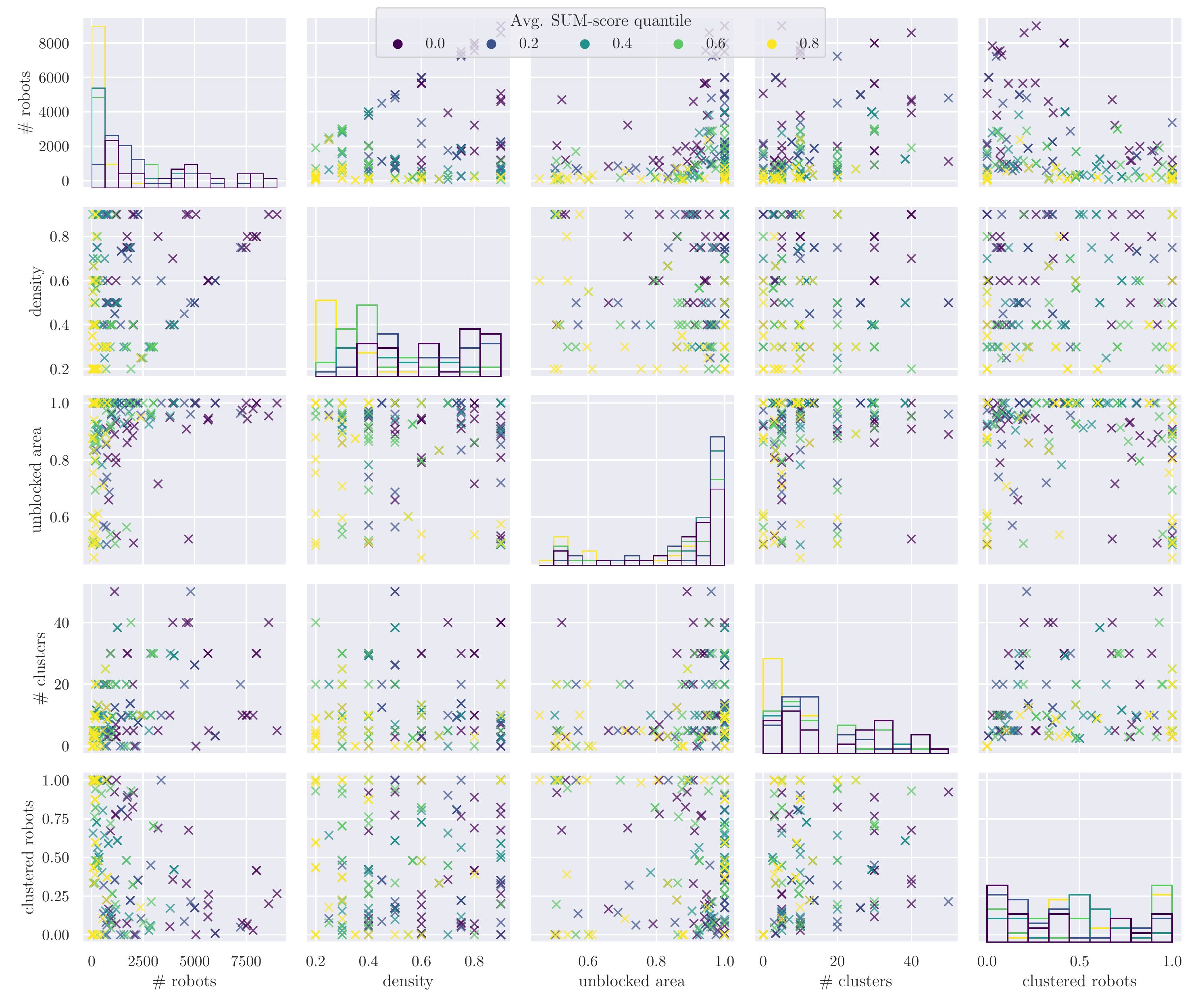}
  \caption{This plot provides some clues for the practical difficulty of instances, indicated
            by low average scores.
	    Because the scores are based on the best submitted solution, this
implies that some teams with special approaches where able to obtain
significantly better results than the ``basic'' approach.  In the plots, we
split the instances into five quantiles, such that the instances with
high average score are shown in yellow, while darker colors indicate lower average score, i.e., instances considered to be difficult.
            The most dominant factor appears to be the number of robots.
            Because it correlates with many other parameters and the distribution is not fully homogeneous, considering individual parameters would be highly skewed.
	    We therefore also show combinations of parameters, especially with
the number of robots to counteract the effect that some parameter ranges have
on average significantly fewer or more robots.  The diagonals show the
distribution of individual parameters; we can see that the easiest
instances appear to be the small ones, because the better quantiles are 
nearly full. 
            Scatter plots show combinations of two parameters, with every instance shown as one point in the plot.
            We can also see that increasing the density without changing the number of robots makes the instances more difficult.
            Moreover, instances without obstacles seem to be easier to solve.
            Due to lack of data, not much can be said for clusters, with difficulty mostly correlating with the number of robots in the available data points.
            }
\label{fig:hardness_SUM}
  %\caption{The number of robots is the most dominant factor when considering the difficulty, here measured in average score (1.0 meaning all teams had the same solution and 0.5 meaning that the average solution had only half the score of the best submitted solution). 
  %Because the parameters are correlated (e.g., more robots allow more clusters), we cannot analyze them individually. Here we can see that denser instance are often hard to solve even for less robots. Additionally, obstacles seem to make the instances slightly more complicated.}
\end{figure}

\begin{figure}[]
  \centering
  \includegraphics[width=0.95\textwidth]{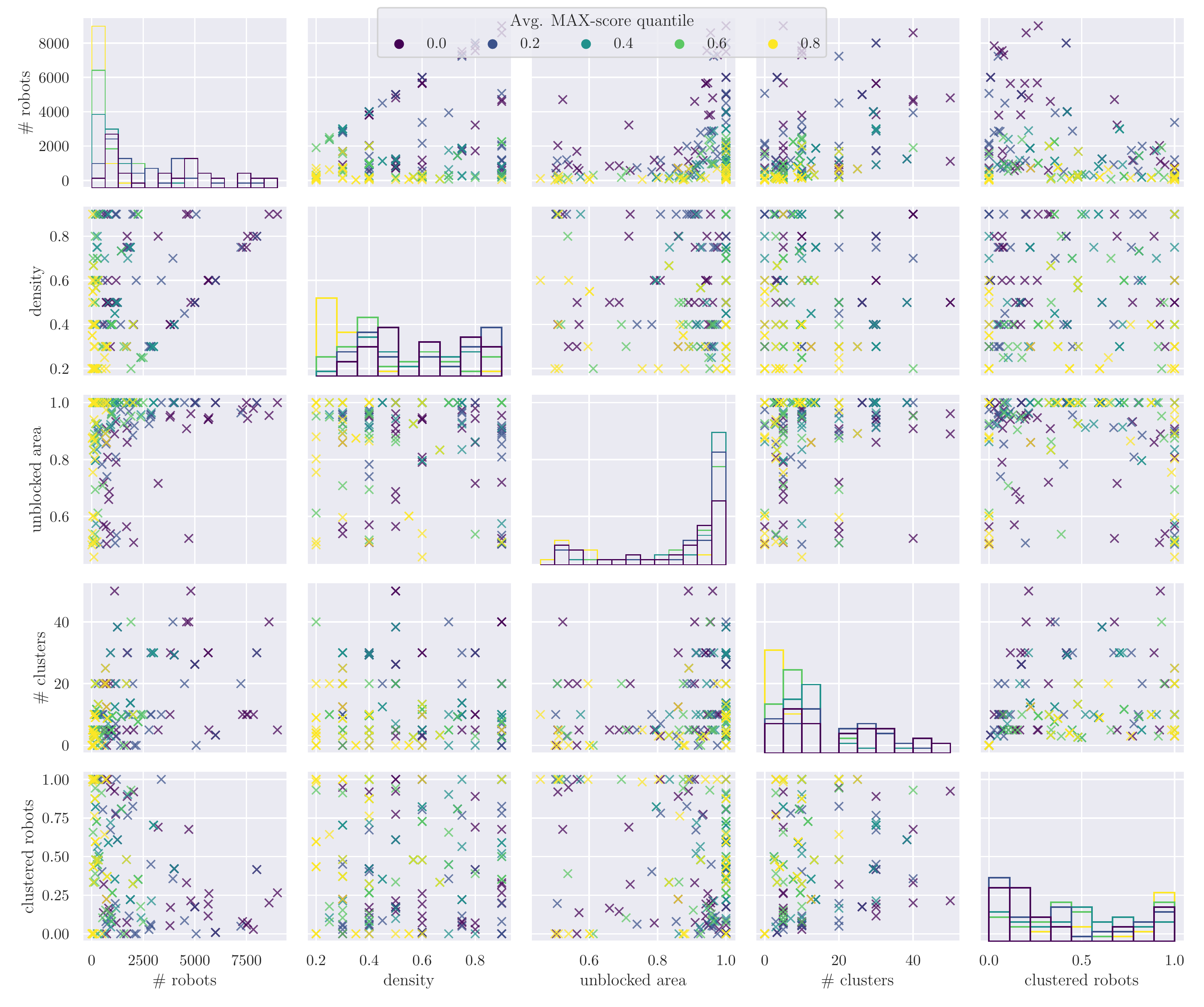}
  \caption{A detailed analysis for MAX instead of SUM; refer to Figure~\ref{fig:hardness_SUM} for the breakdown into subfigures.}
\label{fig:hardness_MAX}
\end{figure}

\section{Conclusions}
The 2021 CG:SHOP Challenge motivated a considerable number of teams to engage in intensive optimization studies.
The outcomes promise further insight into the underlying, important optimization problem. Moreover, the 
considerable participation of junior teams indicates that the Challenge itself motivates a considerable number
students and young researchers to work on practical algorithmic problems.

\bibliography{refs,bibliography}
\end{document}